# Fractal-Based Exponential Distribution of Urban Density and Self-Affine Fractal Forms of Cities

Yanguang Chen, Jian Feng

(Department of Geography, College of Urban and Environmental Sciences, Peking University, Beijing 100871, P.R. China. Email: chenyg@pku.edu.cn)

**Abstract**: Urban population density always follows the exponential distribution and can be described with Clark's model. Because of this, the spatial distribution of urban population used to be regarded as non-fractal pattern. However, Clark's model differs from the exponential function in mathematics because that urban population is distributed on the fractal support of landform and land-use form. By using mathematical transform and empirical evidence, we argue that there are self-affine scaling relations and local power laws behind the exponential distribution of urban density. The scale parameter of Clark's model indicating the characteristic radius of cities is not a real constant, but depends on the urban field we defined. So the exponential model suggests local fractal structure with two kinds of fractal parameters. The parameters can be used to characterize urban space filling, spatial correlation, self-affine properties, and self-organized evolution. The case study of the city of Hangzhou, China, is employed to verify the theoretical inference. Based on the empirical analysis, a three-ring model of cities is presented and a city is conceptually divided into three layers from core to periphery. The scaling region and non-scaling region appear alternately in the city. This model may be helpful for future urban studies and city planning.

# 1 Introduction

The mathematical models on urban density are special spatial correlation functions, which can be used to analyze spatial autocorrelation and spatio-temporal evolution of cities. Urban density distributions fall into two classes--the scale-free distribution without characteristic length, and the scale-dependent distribution with characteristic length. The former indicates the power-law distribution, while the latter mainly include the exponential distribution and the normal distribution. The power-law distribution usually suggests fractal structure, and the fractal dimension can be estimated with the number-radius scaling or box-counting method (Batty and Longley, 1994; Chen, 2012; Frankhauser, 1998; Shen, 2002). The common exponential distribution and normal distribution are not of fractal pattern, suggesting no fractional dimension. In practice, three kinds of distributions are always modeled by three functions: power function, exponential function, and normal function (Gaussian function).

Since Clark (1951) employed the negative exponential function to describe the population density, more than eleven functions have been introduced to characterize urban density (Batty and Longley, 1994; Cadwallader, 1996; Chen, 2010a; McDonald, 1989; Zielinski, 1979). Among all these density models, three ones came to front successively. The first is the negative exponential function known as Clark's model, the second is the Gaussian function known as Sherratt-Tanner's model (Sherratt, 1960; Tanner, 1961), and the third is the inverse power function known as Smeed's model (Smeed, 1963). The Clark model is well-known for geographers, representing the most influential form for urban density. The Sherratt-Tanner model has the advantage of Clark's model because of its simpler expression for mathematical analysis (Dacey, 1970). However, in empirical analysis of urban form, the negative exponential function gains an evident advantage over the Gaussian function. In terms of fractal concepts, Batty and Kim (1992) argued forcibly that the use of Clark's model is fundamentally flawed due to its absence of a parameter indicating space-filling competence of systems. They suggested that the most appropriate form for urban population density models is the inverse power function associated with fractal distribution rather than the negative exponential function indicative of the distribution with characteristic scale.

This paper will present a viewpoint differing to some extent from Batty and Kim (1992). We argue for the suggestion that the inverse power function is very significant for us to study urban density, but we argue against the opinion that Clark's model does not imply fractal structure and space filling. The inverse power function can be used to describe urban density of transport network, while the negative exponential function is suitable for characterizing urban population density in many cases. Therefore, the Clark model cannot always give place to the Smeed model. This paper will reinterpret Clark's model with the ideas from fractals. I will argue that the Clark model is not a simple exponential function, but the one indicating special exponential distribution based on fractal supports and denoting fractal form. Also this paper will reinterpret the claim of Parr (1985a, 1985b), who suggested that the negative exponential function is more appropriate for describing population density in the urban area itself, while the inverse power function is more appropriate to the urban fringe and hinterland (see also Batty and Longley, 1994).

The innovation of the paper lies in following aspects. First, it will show how to understanding the self-affine fractal feature behind the exponential distribution by means of scaling analysis. Second, it will illuminate how to recognize the dimension of the urban density distributions with superficial characteristic scales. This is helpful for us to comprehend the nature of urban space. Third, based on the special exponential distribution with fractal properties, a three-ring city model is proposed for geographers and planners to grasp the spatial structure of cities. The three urban density models mentioned above oppose each other but also complement one another. This paper is devoted to researching into the exponential distribution, and at the same time, this work also discusses the normal distribution and power-law distribution for reference. The exponential model and the normal one can be formally unified in mathematical expression.

## 2. The dimension of the urban distribution with scale

### 2.1 Basic postulates, concepts and analytical methods

For simplicity, we only consider the monocentric cities with single core of growth. In many cases, a polycentric city can be treated as a monocentric one by changing coarse-graining level. Two postulates are put forward as follows. First, the *landform* is a fractal body, which influences urban land use and population distribution. Second, the urban *land use form* is a fractal pattern, and there exists an interaction between population distribution and land use structure. Both the landform and land use form compose the *physical infrastructure* of population distribution. The human aggregation is determined by the physical infrastructure and in turn reacts on it. The physical infrastructure can be regarded as a *fractal support*, on which a city grows and evolves.

If urban density satisfies the exponential distribution or normal distribution, it possesses a parameter indicating *characteristic length*. For example, in Clark's model, the relative rate at which the effect of distance attenuates used to be looked upon as this kind of parameter. The reciprocal of the *rate parameter* is a *scale parameter* indicating the *characteristic radius* ($r_0$) of urban population distribution (Takayasu, 1990). It suggests some mean distance of human activities. In theory, the urban form which is similar to the "fractal dust" in appearance has no distinct boundary (Thomas, 2007; Thomas, 2008). We can identify an urban boundary by a fractal approach (Tannier *et al*, 2011), or the city clustering algorithm (CCA) (Rozenfeld *et al*, 2008). The boundary forms an *urban envelope* (Longley *et al*, 1991). The region within the envelope can be thought of as the area of a city (*A*). The radius of the circle with the same area as the urban area is termed *boundary radius* ($R_b=(A/\pi)^{1/2}$), representing the distance from a city center to its boundary.

If the characteristic radius of a city is a real constant independent of the city size considered, Clark's model is the conventional exponential function and has no singularity. A speculation is that the characteristic radius varies with the city size defined, and there is a scaling relation under dilation between the characteristic radius ($r_0$) and boundary radius ($R_b$). If so, Clark's model should be regarded as a special exponential function indicating self-affine fractal form, which will be validated in next section. The normal model can be handled in the same way. One of the keystones of this study is to illustrate this kind of scaling relation with observational evidence. The main analytical methods employed by this article include mathematical modeling and empirical analysis. The mathematical methods involve scaling analysis, spatial correlation analysis, and spectral analysis, which can be called "*3S analysis*" of cities. The 3S analysis is very effective to deal with the complicated mathematical models through simple ways.

As prearrangement, three fractal concepts should be made clear here. The first is *real fractal* (R-fractal), which needs no special explanation (see Batty and Longley, 1994; Frankhauser, 1994; Mandelbrot, 1983). The second is *pseudofractal* (P-fractal), which suggests the fractional dimension coming from the non-fractal systems. This can be regarded as "fractal rabbits" (Kaye, 1989). The pseudofractals are always generated by the errors resulting from mathematical transformation and approximate treatment. The third is *quasifractal* (Q-fractal), which refers to such a case: intuitionally there is no fractal dimension, but empirically come out a fractional dimension that cannot be strictly distinguished from the real fractal dimension in practice. We have several approaches to demonstrating that the dimension of the common exponential distribution and normal distribution can be taken as *d*=2. However, for the special exponential distribution and normal distribution, there exists a local fractal dimension.

## 2.2 Exponential distribution and self-affine fractal form

The special exponential model suggesting latent fractal nature can be derived from Clark's law. For the population density $\rho(r)$ at distance $r$ from the center of the city ($r=0$), the exponential function can be expressed as

$$\rho(r) = \rho_0 e^{-r/r_0}, \tag{1}$$

where $\rho_0$ denotes the proportionality coefficient, which is expected in theory to equal the central density, i.e., $\rho_0=\rho(0)$, and the scale parameter $r_0$ is the characteristic radius of urban population distribution. In urban geography, equation (1) is the equivalent of Clark's model. Given a city radius $R_b$, it follows that $r_0$ is a constant, and the scales of parameters and variable are as below: $0 \le r \le R_b$, $0 < r_0 < R_b$, $r_0 < R_b < \infty$. Integrating equation (1) over $r$ by parts yield a cumulative distribution

$$P(r) = 2\pi \int_0^r x\rho(x)dx = 2\pi r_0^2 \rho_0 [1 - (1 + \frac{r}{r_0})e^{-r/r_0}], \tag{2}$$

in which $x$ is a distance ranging from 0 to $r$, $P(r)$ denotes the cumulative population within a radius of $r$ of the city center. Thus, according to l'Hospital's rule (or Bernoulli's rule), if $R_b$ is large enough, the total population of the city ($P_T$) will be given by

$$P_T = P(R_b) = 2\pi r_0^2 \rho_0 [1 - (1 + \frac{R_b}{r_0})e^{-R_b/r_0}] \approx 2\pi r_0^2 \rho_0, \tag{3}$$

which can be derived by using the method of entropy maximization (Chen, 2008).

A local scaling relation can be derived from equation (2) with the Taylor series expansion. The series expansion is an effective approach to transforming a nonlinear system into a linear structure. In fractal studies, Taylor's series is often employed to make a local scaling analysis (Turcotte, 1997). When $r<r_0$, equation (2) can be expanded into a Taylor series, and approximately, we have

$$P(r) \approx 2\pi r_0^2 \rho_0 [1 - (1 + \frac{r}{r_0})(1 - \frac{r}{r_0})] = 2\pi r_0^2 \rho_0 (\frac{r}{r_0})^2 = 2\pi \rho_0 r^d, \tag{4}$$

where $d=2$ in theory but we have $d<2$ in practice. This is the first scaling equation that we need. Generally, $d$ value varies from 1.6 to 1.9. The value of $d$ can be regarded as the dimension of the phenomena following the exponential distribution. Since $d$ is valid only for the central area ($r<r_0$), it can be treated as a *local dimension*, and termed *inner dimension* of urban population.

Suppose that there exists a fractal support, on which all the urban inhabitants are distributed.

For given scale $R_b$, population density follows Clark's law. If urban form looks like a random fractal dust in two dimensions, we cannot find a certain city radius. For the simplest case, $r_0$ is considered to be a constant. Then equation (2) conforms to a self-affine scaling relation

$$P(\lambda r, \lambda r_0) = 2\pi(\lambda r_0)^2 \rho_0 [1 - (1 + \frac{\lambda r}{\lambda r_0})e^{-\lambda r/\lambda r_0}] = \lambda^2 P(r, r_0), \tag{5}$$

in which $\lambda$ is a scale factor. This self-affine transformation is defined in the Euclidean framework. In urban studies, the parameter $R_b$ and $r_0$ represent different measurements. Two measurements of the same system always comply with the allometric scaling law (Chen and Jiang, 2009). Thus, for a city on fractal supports, we assume

$$r_0 \propto R_b^q, \tag{6}$$

where $q$ is a scaling exponent. This implies that if we change $R_b$ value, the characteristic radius $r_0$ will change accordingly. If $R_b$ is large enough, in light of equation (3), we will obtain

$$P_T(R_b) \approx 2\pi \rho_0 r_0^2 \propto R_b^{2q} = R_b^D, \tag{7}$$

where $D=2q$ is the fractal dimension of population distribution in a large city region. This is the second scaling equation that we need, and it can be used to explain Parr's viewpoint (Parr, 1985a; Parr, 1985b). Since $D$ is valid only for a great scale ($R_b > R_c$, $R_c$ is a critical value), it should be treated as another local dimension for urban population, *outer dimension*. The local dimension value comes in between 1 and 2 ($1 \leq D \leq 2$). Equation (6) suggests

$$r_0(\lambda R_b) = \lambda^q r_0(R_b). \tag{8}$$

For the scaling analysis, the radius $R_b$ has no essential difference in value from $r$, thus equation (8) can be formally replaced by

$$r_0(\lambda r) = \lambda^q r_0(r). \tag{9}$$

Then, combining equations (2) and (9), we have a general self-affine scaling relation such as

$$P(\lambda r, \lambda^{1-q} r_0) = 2\pi(\lambda^{1-q} r_0)^2 \rho_0 [1 - (1 + \frac{\lambda r}{\lambda^{1-q} r_0(\lambda r)})e^{-\lambda r/(\lambda^{1-q} r_0(\lambda r))}]$$
$$= \lambda^{2(1-q)} P(r, r_0) = \lambda^{d-D} P(r, r_0) \tag{10}$$

where $d=2$ is the inner dimension, and $D=2q$ the outer dimension. This self-affine transformation is defined in a fractal framework. Therefore, the scaling exponent can be expressed as $a=d-D$. In order to keep the scaling relation for the exponential function, we need two scale factors ($\lambda$, $\lambda^{1-q}$).

This is just the character of self-affinity. According to equation (8), if $r_0$ is independent of $R_b$ and thus $r$, we will have $q=0$. In this instance, the outer dimension vanishes, and we have a *global dimension d=2*. This is a Euclidean dimension.

## 2.3 Normal distribution and fractal structure

Understanding the exponential model is instructive for us to comprehend the Gaussian diffusion model (GDM). The normal distribution function can be written as

$$\rho(r) = \rho_0 e^{-r^2/2r_0^2}, \quad (11)$$

which was employed by Sherratt (1960) and Tanner (1961) to describe city population density. Integrating equation (11) gives the Weibull distribution

$$P(r) = 2\pi \int_0^r x\rho(x)\mathrm{d}x = 2\pi r_0^2 \rho_0 (1 - e^{-r^2/2r_0^2}), \quad (12)$$

where the symbols fulfill the same roles as in equations (1) and (2). Obviously, if $r$ is large enough, we will have $P(r) \approx 2\pi r_0^2 \rho_0 = P_T$, which refers to the total population of a city. For $r<r_0$, expanding equation (12) into a Taylor series yield an approximate relation such as

$$P(r) \approx 2\pi r_0^2 \rho_0 \{1 - [1 - (\frac{r}{\sqrt{2}r_0})^2]\} = \pi r_0^2 \rho_0 (\frac{r}{r_0})^2 = \pi \rho_0 r^d, \quad (13)$$

where $d$ is the inner dimension. Theoretically, the dimension is about $d=2$, but empirically we have $d<2$. Generally, $d$ value comes between 1.8 and 2.0. Similar to the exponential cumulative distribution, the Welbull distribution follows the self-affine scaling law

$$P(\lambda r, \lambda^{1-q}r_0) = 2\pi(\lambda^{1-q}r_0)_0^2 \rho_0[1 - e^{-(\lambda r)^2/2(\lambda^{1-q}r_0(\lambda r))^2}] = \lambda^{d-D}P(r,r_0). \quad (14)$$

which also gives a scaling exponent such as $a=d-D$, where $d=2$ is the local dimension, and $D=2q$ the outer dimension.

The exponential model, equation (1), and the normal model, equation (11), can be formally unified. According to equation (6) and equation (9), we have $r_0 \propto r^q$. Substituting this relation into equation (1) gives

$$\rho(r) = \rho_0 e^{-r/kr^q} = \rho_0 e^{-r^{1-q}/k} = \rho_0 e^{-r^\sigma/\sigma_0'}, \quad (15)$$

where $k=\sigma r_0'$ is a proportionality coefficient, $\sigma=1-q$ is an exponent, and $r_0'$ is the rescaled characteristic radius. If $\sigma=1$, then equation (15) will go back to Clark's model; if $\sigma=2$, equation (15) will turn to Sherratt-Tanner's model; If $\sigma=1/2$, equation (15) will become Parr's model (Parr, 1985a). Generally, the exponent value varies from 0 to 2 (Chen, 2010a). In this context, the exponential distribution and normal distribution are two special cases. However, it is difficult to make mathematical analysis based on equation (15), but it is easy to implement scaling analysis based on equations (1) and (11). In scaling analysis, different variables can be separately treated. The key scaling relation of this paper is equation (6), which cannot be theoretically demonstrated, but can be empirically tested and verified. If equation (6) is confirmed by empirical facts, Clark's model will be revised with the idea from the self-affine fractals.

# 3 Empirical evidence: the case of Hangzhou

## 3.1 Estimation of the inner dimension

The datasets of Hangzhou, China, will be employed to testify the self-affine fractal model of urban population density. First, we will show how to estimate the inner dimension and outer dimension. Second, the dimension values will be used to explain the spatial evolution of Hangzhou. Third, in particular, the scaling relation between the boundary radius ($R_b$) and characteristic radius ($r_0$) will be verified, and thus the self-affine property of the special exponential distribution will be validated by the observational data.

Hangzhou is selected for our empirical analyses because Clark's model has already been applied to its population density data (Feng, 2002). Four sets of census data of the city in 1964, 1982, 1990, and 2000 are available. The census enumeration data is based on *jie-dao*, or sub-district (Wang and Zhou, 1999), which bears an analogy with urban zones in Western literature (Batty and Kim, 1992). In fact, a zone or sub-district (*jie-dao*) is an administrative unit comprising several city blocks defined by streets and other physical features. The study area is confined in the combination of city proper and its outskirts, and this scope, approximately, comes between the urbanized area (UA) and the metropolitan area (MA) of Hangzhou. The zone with maximum population density is defined as the center of the city, and the data are processed by means of spatial weighed average based on concentric circles (Chen, 2008). The method of data processing is illuminated in detail by Feng (2002). The length of sample path is 26, and the maximum urban radius is 15.3 kilometers. The radii of the concentric circles range from 0.3 to 15.3 km, and the sampling step-length is 0.6 km (Feng, 2002; Feng and Zhou, 2005).

If we fit the density data to the inverse power function, i.e., the Smeed model, two problems will arise. First, the data points do not match well with the trend line. Second, the scaling exponent is greater than 1. This suggests that the fractal dimension is great than 2. In other words, the value of the scaling exponent is hard to be geometrically interpreted. In contrast, if we fit the data to the negative exponent function, equation (1), the results are acceptable in the statistical sense. The cumulative population within the radius of $r$ can be fitted to equation (2). Taking the data set in 2000 as an example, we have the following result

$$P(r) = 2675611.039[1-(1+\frac{r}{3.3494})\exp(-\frac{r}{3.3494})].$$

The goodness of fit is about $R^2=0.9962$ (Figure 1). The characteristic radius is estimated as $r_0=3.3494$ (accordingly, Clark's model gives $r_0 \approx 3.9463$), and the total population of the city is estimated as $P_T=2,675,611$ (the result from the census data is $P_T=2,451,319$).

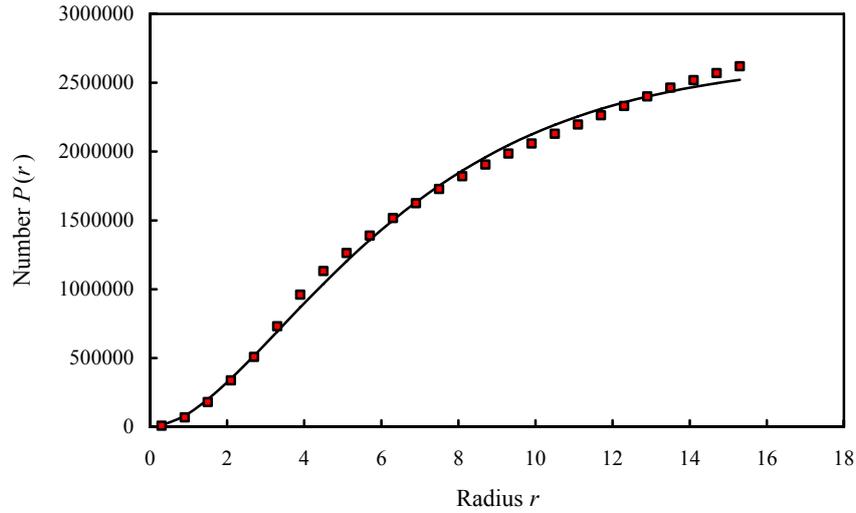

**Figure 1 The relation between radius and corresponding population of Hangzhou (2000)**

By means of equation (1), we can evaluate the characteristic radius $r_0$ based on the whole data in every year. Taking $r_0$ value as division, we can divide the cumulative population in each year into two scale ranges: the range of inner zone ($r<r_0$) and the range of outer zone ($r \geq r_0$). In both the scale ranges, the cumulative population follows the power law approximately (Figure 2). Fitting the data within the inner range ($r<r_0$) to equation (4) yields the values of the inner dimension. The scaling exponent values of the outer zones are also displayed for comparison (Table 1). By the way, the character of curves displayed by Figure 2 is very similar to that of what is called bi-fractals (White and Engelen, 1994). Bi-fractal curves are also associated with self-affinity, and maybe they are associated to a degree with the special exponential distribution.

**Table 1 The inner dimension values and the related parameters of Hangzhou's population distribution**

| Zone | Parameter name | Parameter value | | | |
|---|---|---|---|---|---|
| | | 1964 | 1982 | 1990 | 2000 |
| Inter zone | Inner dimension ($d$) | 1.7675 | 1.7554 | 1.8077 | 1.8709 |
| | Goodness of fit ($R^2$) | 0.9988 | 0.9990 | 0.9986 | 0.9991 |
| | Division radius ($r_c$) | 3.3 | 3.3 | 3.3 | 3.9 |
| Division | Characteristic radius ($r_0^*$) | 3.5638 | 3.6711 | 3.6284 | 3.9463 |
| Outer zone | Scaling exponent ($b$) | 0.5554 | 0.5848 | 0.5818 | 0.6675 |
| | Goodness of fit ($R^2$) | 0.9990 | 0.9988 | 0.9953 | 0.9913 |

**Note**: The characteristic radius ($r_0^*$) is estimated with Clark's model based on the whole data ($0<r\leq15.3$km).

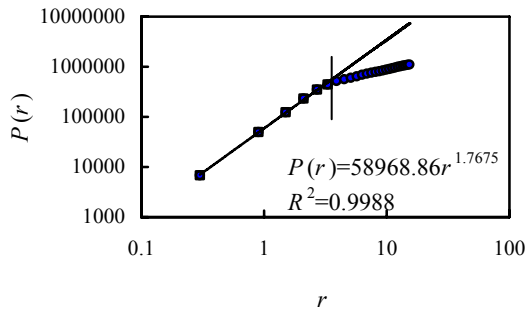

a. 1964

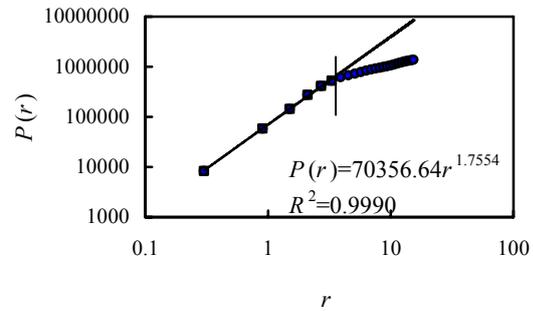

b. 1982

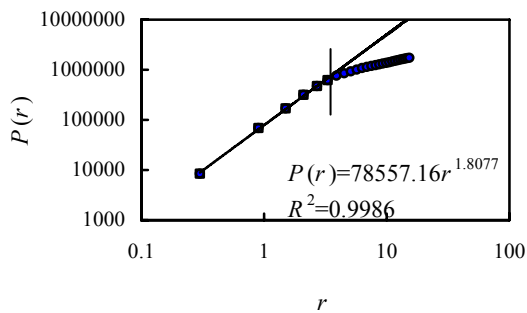

c. 1990

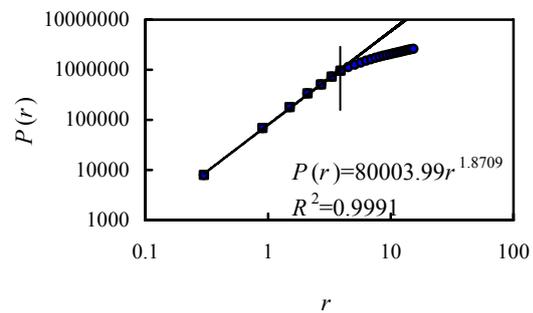

d. 2000

**Figure 2 The local fitting of the cumulative population data of Hangzhou to the power function**

[**Note**: The vertical line in each subplot indicates the division between the different scaling ranges.]

## 3.2 Estimation of the outer dimension

One of the key points in this study is to illustrate the scaling ration between the characteristic radius ($r_0$) and the boundary radius ($R_b$). Based on the scaling relation, we can revise Clark's model and thus estimate the outer dimension of the population distribution. Hangzhou's urban form has been demonstrated to be fractal by box-counting method (Feng and Chen, 2010). As a fractal, it is hard to find the urban boundary objectively. Thus, we can subjectively define a radius as the equivalent radius of urban area. Changing the scope of urban field, we can obtain different urban area, and thus different boundary radius ($R_b$). The computation results show that the characteristic radius ($r_0$) is assuredly a variable rather than constant. Its value depends on the boundary radius. The characteristic radius will become shorter if we increase the boundary radius. However, if the boundary radius is greater than some value ($R_t$), the trend will be suddenly reversed, and the characteristic radius will become longer with increasing the boundary radius (Figure 3).

A difficult problem is how to find the critical value of boundary radius, which will be employed to define the scaling range of the outer dimension. The critical radius ($R_c$) is different from the extremum of the boundary radius ($R_t$). The former indicates the starting point of the scaling range, while the latter suggests the minimum characteristic radius and the turning point of the characteristic radius changing with the boundary radius. The standard of determining the critical radius is as follows. First, the critical radius must be ascending with the boundary radius, i.e., $R_c \geq R_t$. Second, the goodness of fit of the data to the power function must be high enough. Third, the scaling exponent must be rational, that is, $0 < q \leq 1$. The critical value can also be used as a division point, by which the urban field is divided into two parts: inner part and outer part.

The analytical results of outer part show that there exists a scaling relation between the characteristic radius and the boundary radius when $R_b \geq R_c$ (Figure 4). The scaling exponent ($q$), the corresponding outer dimension ($D$), and some related variables and statistics are tabulated as follows (Table 2). For the inner part, there also is weak scaling relation, but the scaling exponent is a negative. The results are listed in the same table for comparison.

**Table 2 The outer dimension values and the related parameters and variables of Hangzhou's population distribution**

| Zone | Parameter or variable | Parameter/variable value | | | |
|---|---|---|---|---|---|
| | | 1964 | 1982 | 1990 | 2000 |
| Inter zone | Scaling exponent ($\sigma'$) | -0.5996 | -0.6176 | -0.5848 | -0.6148 |
| | Goodness of fit ($R^2$) | 0.7756 | 0.8372 | 0.8853 | 0.9205 |
| | Scaling range ($r$) | 2.1-8.1 | 2.1-7.5 | 2.1-8.7 | 0.3-10.5 |
| Division | Critical radius ($R_c$) | 8.7 | 8.1 | 9.3 | 11.1 |
| Outer zone | Scaling exponent ($\sigma$) | 0.8207 | 0.8328 | 0.7607 | 0.5161 |
| | Goodness of fit ($R^2$) | 0.9966 | 0.9898 | 0.9955 | 0.9964 |
| | Scaling range ($R_b$) | 8.7-15.3 | 8.1-15.3 | 9.3-15.3 | 11.1-15.3 |
| | Outer dimension ($D$) | 1.6414 | 1.6656 | 1.5214 | 1.0322 |
| | Scaling exponent ($a=2-D$) | 0.3586 | 0.3344 | 0.4786 | 0.9678 |

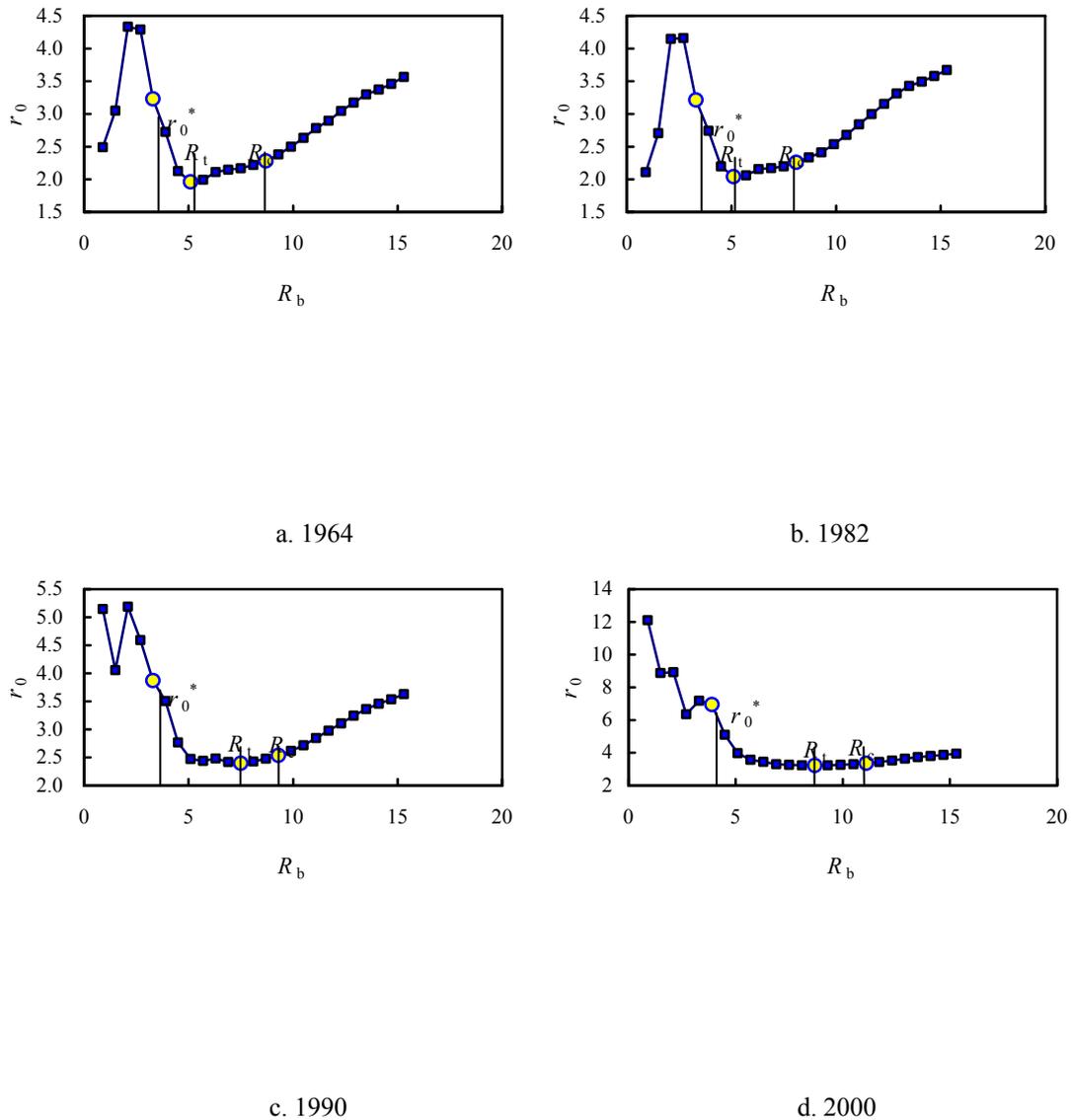

a. 1964                      b. 1982

c. 1990                      d. 2000

**Figure 3 The characteristic radius changes with the boundary radius of Hangzhou**

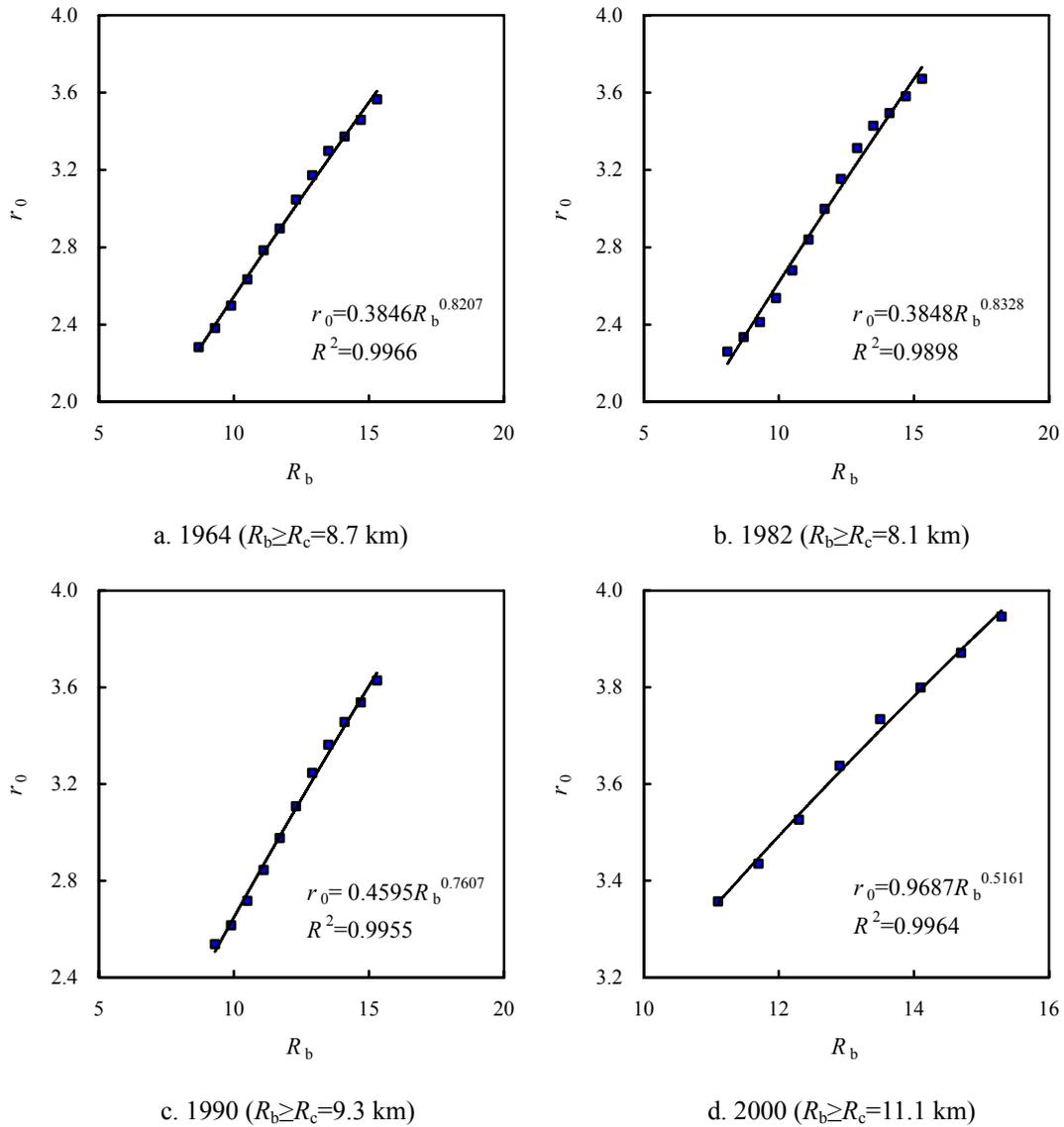

a. 1964 ($R_b \geq R_c$=8.7 km)  b. 1982 ($R_b \geq R_c$=8.1 km)

c. 1990 ($R_b \geq R_c$=9.3 km)  d. 2000 ($R_b \geq R_c$=11.1 km)

**Figure 4 The scaling relation between the boundary radius and the characteristic radius of Hangzhou ($R_c \leq R_b \leq$15.3km)**

[**Note**: The goodness of fit of power function is greater than those of linear function, exponential function, and logarithmic function for 1964, 1982, and 1990. In 2000, the goodness of fit of logarithmic function is $R^2$=0.9972, which is greater than that of power function.]

The case of Hangzhou gives support to the hypothesis that the value of a city's characteristic radius ($r_0$) scales itself to vary with the city's boundary radius ($R_b$). Within certain scale range, the scaling relation indicating self-similarity, equation (8), comes into existence, and thereby the scaling relation indicating self-affinity, equation (10), is empirically confirmed. Both the inner dimension and outer dimension belong to what is called "radical dimension" (Frankhauser and Sadler, 1991). Now, let's make a simple analysis of the spatio-temporal evolution of Hangzhou with the radical dimension values and related variables. First, the inner dimension (*d*) went ascending, while the outer dimension (*D*) descended in the mass from 1964 to 2000 (Figure 5). This suggests that the population distribution of the city tended to concentrating into the city proper. Second, the critical radius ($R_c$) became longer and longer as a whole (Table 2). This implies a process of urban growth along with the population aggregation. Third, the data in 1982 make an exception. Compared with the values in 1964, the inner dimension went down, the outer dimension went up, and the critical radius became short. These are due to the Great Proletarian Cultural Revolution (1966-1976), vulgarly termed "the 10-year man-made catastrophe", which resulted in Hangzhou's stagnation of development. To sum up, the dimension change trend argues for the viewpoint of population concentration (Chen and Jiang, 2009), but unfortunately, to some extent, against the opinion that Hangzhou evolved into the stage of suburbanization (Feng, 2002).

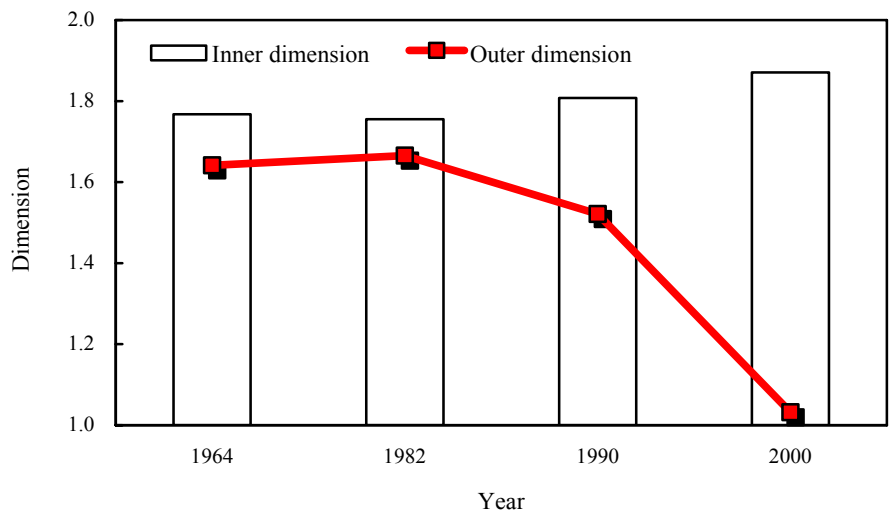

**Figure 5 The changing trend of the inner dimension and outer dimension of Hangzhou**

## 4 Questions and discussions

### 4.1 Three-ring city models

The relationships between urban morphology and size had been a much neglected realm in spatial analysis before fractal theory was introduced to urban studies, and only since the renaissance prompted by the ideas from fractal geometry has the interest of geographers been rekindled in these questions (Batty, 2008; Longley *et al*, 1991). The focus of this paper is not the empirical study, but the theory and mathematical modeling of cities. However, the case study lends empirical support to the author's theory. Further, simulation experiments can help us understand both the urban density models and urban evolution. Generally, the computer simulation of cities is based on two kinds of spatial dynamics: Brown motion (random walk) and fractional Brownian motion (fBm). The Brownian-motion-based simulation fall into two classes: one is random diffusion, and the other is diffusion-limited aggregation (DLA). In urban studies, DLA has been employed by Batty (1991), Batty and Longley (1994), and Fotheringham *et al* (1989) to simulate urban growth and model urban form.

From the angle of view of cumulative distribution, the process of urban growth is similar to DLA to some extent (see Appendix). If we use the box-counting method to make a measurement, both urban form and DLA are fractals; however, if we apply the area-radius scaling to urban form or DLA, neither of them follows the power law strictly. A comparison between the DLA, GDM, and urban growth is instructive for us to research deeply into cities. The Gaussian random diffusion process is based on Brownian motion without spatial correlation. Its density distribution is normal and can be described with Tanner-Sherratt's model. DLA is also based on Brownian motion but it has local correlation. The radial density distributions of the DLA patterns are expected to be power-law distribution but really linear distribution, exponential distribution or local power-law distribution. Its density model is not clear. Urban growth seems to be based on fBm rather than Brownian motion. Its density distributions take on power-law distribution or exponential distribution. Urban population growth bears more analogy to the DLA than that to the GDM (Longley and Mesev, 1997). This suggests that a city evolves through local spatial correlation, and the normal distribution function is not very proper to model urban density.

The DLA model has clear limitations for our understanding urban evolution. Makse *et al* once (1995, page 608) pointed out: "The DLA model predicts that there should exist only one large fractal cluster, which is almost perfectly screened from incoming 'development units' (…), so that almost all of the cluster growth takes place at the tips of the cluster's branches." In real urban areas, however, development attracts further development, and different development units are correlated instead of being added to cluster at random. In view of these facts, Makse *et al* (1995, 1998) proposed an alternative model, which corresponds to the correlated percolation model (CPM) in the presence of a density gradient, to reproduce the morphology of cities and describe urban growth dynamics. The CPM offers the possibility of predicting the global properties such as scaling behavior of urban form and growth. The works of Makse *et al* (1995, 1998) are based on urban land use form rather than urban population distribution, and the CPM seems to be more suitable for polycentric cities rather than monocentric cities. Recently, Rozenfeld *et al* (2008, 2011) demonstrated the existence of long-range spatial correlations in population growth. The real urban growth patterns may come between the CPM and DLA. It is interesting and revealing for us to draw a comparison between CPM, DLA, GDM, and urban growth (Table 3).

Table 3 Comparison between the random diffusion, DLA, CPM, and urban growth

| Type | Model | Evolution | Density distribution | Micro motion | Macro pattern | Urban pattern |
|---|---|---|---|---|---|---|
| Simulation experiments | GDM | Random diffusion | Normal | Brownian motion | Random | Not Available |
| | DLA | Random aggregation and diffusion | Local power-law, exponential, or linear | Brownian motion | Fractal | Population |
| | CPM | Correlated percolation | Power-law | fBm | Fractal | Land use form |
| Observation | Urban growth | Random aggregation and diffusion | Power-law or exponential | fBm | Fractal or Q-fractal | Morphology |

The normal distribution indicates simple structure, while the power-law distribution implies complex structure (Goldenfeld and Kadanoff, 1999). The exponential distribution seems to suggest the structure coming between the simple and the complex. If the distributions with characteristic scales are based on the fractal supports, both the exponential distribution and normal distribution will possess complex structure associated with self-affine fractals. In terms of the empirical analysis, a three-ring model of fractal cities associated with the special exponential distribution can be presented here. The inner layer is a scale-free region, and the cumulative distribution can be modeled with a power function. This area has fractal feature or quasi-fractal nature. The middle layer is a non-scaling region, and the density distribution is with characteristic scale. This belt can be treated as exponential distribution. The outer layer is also a scaling region, and the density distribution is of fractal pattern (Figure 6(a)). As a whole, the form is a self-affine fractal with two scale factors. For example, for Hangzhou in 1964, the scale of radius is as follows: the inner layer is within the radius of $r_0^* \approx 3.6$ km, the outer layer is outside the radius of $R_c \approx 8.7$ km, and the middle layer comes between the radii of $r_0^*$ and $R_c$, i.e., $3.6 < r \leq 8.7$ km. The rest may be deduced by analogy (Figure 3, Tables 1 and 2). The outer layer is scale-free region, which makes a new annotation for the suggestion of Parr (1985a, 1985b), who argued that the density of the urban fringe and hinterland can be modeled with inverse power functions.

**Table 4 Comparison of fractal properties between the power-law distribution and the special exponential distribution**

| Distribution | The first scale range | The second scale range | The third scale range | Fractal property |
|---|---|---|---|---|
| Power-law distribution | Scale-dependent range | Scaling range; Scale-free region | Scale-dependent range | Self-similar fractal |
| Exponential distribution | Scaling range; Scale-free region | Scale-dependent range; Memory-free region | Scaling range; Scale-free region | Self-affine fractal |

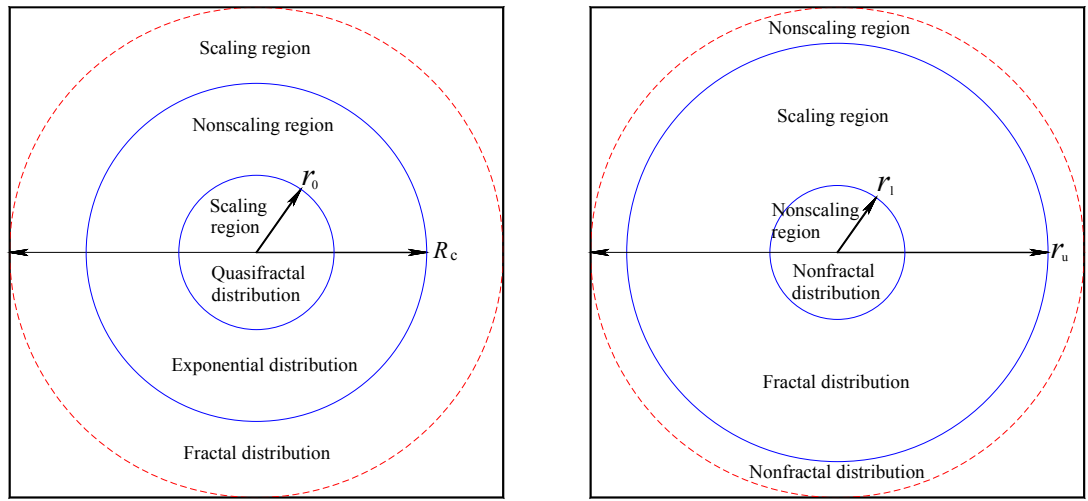

a. Based on exponential distribution    b. Based on power-law distribution

**Figure 6 Two sketch maps of the three-ring models of fractal cities**

With regard to the power-law distribution, we can build another three-ring model of fractal cities, which forms a striking contrast to the exponential function based model (Table 4). The inner part and the outer part are non-scaling regions without fractal properties. The middle part is a scaling range indicating self-similar fractal structure (Figure 6(b)). The power function of density distribution can be demonatrated to be a special correlation function, and the radial dimension is in fact a zero-order correlation dimension. When distance is too long or too short, the spatial correlation is always disabled or out of order (Chen, 2010b). Therefore, the correlation dimension is generally valid within certain range of scale, which suggests a scale-free region. The power-law distribution of urban density will be specially discussed in a companion paper. This article is devoted to the exponential distribution, with the normal distribution as a contrast.

**4.2 The dimension of the distributions with characteristic scales**

Spatial analysis is an indispensable approach to urban studies. The key rests with the definitions of geographical space. For the scale-dependent systems, the space can be defined by distance. However, for the scale-free systems, the space should be defined by fractal dimension parameters rather than distance variables. The dimension parameters are as significant as the distance variables for spatial analysis of cities. A pending question is how to determine the dimension values of the scale-dependent systems. The scale-free distribution follows power laws, which suggest that the similar characters appear at various scales. We can estimate the fractal dimension using scaling analysis. In contrast, the exponential and normal distributions are of characteristic scales, suggesting no fractal dimension. However, if this kind of distributions is based on fractal supports, they also have certain fractal parameters, including fractal dimension and quasi-fractal dimension, the latter is sometimes an approximation of Euclidean dimension. What is more, a *global dimension* can be estimated for these distributions. By the spatial correlation analysis and Fourier transform, we can find another approach to understanding the global dimension of the exponential distribution as well as the normal distribution.

The spatial correlation function can be derived from the negative exponential function as follows

$$C(r) = \int_{-\infty}^{\infty} \rho(x)\rho(x+r)\mathrm{d}x = r_0 \rho_0^2 e^{-r/r_0} = r_0 \rho_0 \rho(r), \qquad (16)$$

where $C(r)$ denotes the density-density correlation function. This implies that the correlation function is directly proportional to the density function. Taking $x=0$ in equation (16), which indicates that one point is fixed to the city center, we have a special correlation function—the central correlation function $C_0(r)=\rho_0\rho(r)=C(r)/r_0$. This suggests that the density-density correlation function has no essential difference from the central correlation function, and the central correlation function is equivalent to the density function. Therefore, the exponential function is actually a memory-free function, but for many cases, only the middle part is really "memoryless".

The density of *energy spectrum* can be obtained from the correlation function through the Fourier transform. The result is

$$S(k, r_0) = \int_{-\infty}^{\infty} C(r) e^{-i2\pi kr} dr = r_0 \rho_0^2 \int_{-\infty}^{\infty} e^{-r/r_0} e^{-i2\pi kr} dr, \quad (17)$$

where $S(\cdot)$ denotes the spectral density of "energy", and $k$ is the *wave number* (Chen, 2008). This relation possesses a self-affinity along the two directions, $k$ and $r_0$. Suppose that there exists a scaling relation between $k$ and $r_0$. Given a scale factor $\zeta$, it follows

$$r_0(\zeta k) \propto \zeta^{-p} r_0(k), \quad (18)$$

where $p$ is a scaling exponent ($p \geq 0$). In fact, if equation (8) is validated, equation (18) can be demonstrated to be true due to that $k$ varies as $r$. Then we have such a self-affine scaling relation

$$S(\zeta k, \zeta^{p-1} r_0) = \zeta^{-1} [\zeta^{p-1} r_0(\zeta k)] \rho_0^2 \int_{-\infty}^{\infty} e^{-\zeta r/r_0} e^{-i2\pi k \zeta r} d(\zeta r) = \zeta^{-2} S(k, r_0). \quad (19)$$

The solution to equation (19) is the wave-spectrum relation

$$S(k) \propto k^{-2} = k^{-\beta}. \quad (20)$$

in which $\beta=2$ refers to the spectral exponent. The numerical relation between the spectral exponent and fractal dimension is $\beta=2(D-1)$ (Chen, 2010b). So the global dimension can be proved to be $D=\beta/2+1=2=d$. This implies that the global dimension ($D$) equals the Euclidean dimension ($d$) of the embedding space in which urban form is examined.

For the normal distribution, the density-density correlation function can be derived as

$$C(r) = \int_{-\infty}^{\infty} \rho(x) \rho(x+r) dx = \sqrt{\pi} r_0 \rho_0^2 e^{-r^2/4r_0^2} = r_0 \rho_0 \sqrt{\pi \rho_0 \rho(r)}, \quad (21)$$

The energy-spectral density can be derived from equation (21) through the Fourier transform, and we have

$$S(k, r_0) = \int_{-\infty}^{\infty} C(r) e^{-i2\pi kr} dr = \sqrt{\pi} r_0 \rho_0^2 \int_{-\infty}^{\infty} e^{-r^2/(4r_0^2)} e^{-i2\pi kr} dr. \quad (22)$$

Also this relation possesses a self-affinity along the two directions, $k$ and $r_0$. For the standard case, the scaling relation is

$$S(\zeta k, \zeta^{-1} r_0) = \zeta^{-1} \sqrt{\pi} (r_0/\zeta) \rho_0^2 \int_{-\infty}^{\infty} e^{-(\zeta r)^2/(4r_0^2)} e^{-i2\pi k \zeta r} d(\zeta r) = \zeta^{-2} S(k, r_0), \quad (23)$$

where the notation is the same as in equation (19). We can derive the result similar to equation

(20), which suggests a global dimension $D=d=2$ (Table 5).

The spectral analysis can be used to investigate the global dimension of Hangzhou city. Based on the power-law relation of wave number and spectral density, the scaling exponent values were estimated as follows: $\beta_{1964}=1.4888$, $\beta_{1982}=1.4350$, $\beta_{1990}=1.6637$, and $\beta_{2000}=1.7983$. Thus the global dimension values are about $D_{1964}=1.7444$, $D_{1982}=1.7175$, $D_{1990}=1.8318$, and $D_{2000}=1.8992$. Accordingly, the values of goodness of fit are $R^2_{1964}=0.9246$, $R^2_{1982}=0.9195$, $R^2_{1990}=0.9655$, and $R^2_{2000}=0.9494$, respectively. As a whole, the Hangzhou's population distribution took on the fractal property. The global dimension value is closer and closer to the $d=2$ over time.

Table 5 Comparison between different dimensions of the distributions with characteristic scales

| Condition | Distribution | Distribution | Theoretical value | Experimental value |
|---|---|---|---|---|
| $r_0$=constant | Exponential | Global | 2 | 1.6~2 |
| | | Local | 2 | 1.6~2 |
| | Normal | Global | 2 | 1.7~2 |
| | | Local | 2 | 1.7~2 |
| $r_0$ scaling with $R_b$ | Exponential | Inner | 2 | 1.6~2 |
| | | Outer | 1~2 | 1~2 |
| | | Global | 2 | 1.5~2 |
| | Normal | Inner | 2 | 1.7~2 |
| | | Outer | 1~2 | 1~2 |
| | | Global | 2 | 1.5~2 |

## 5 Conclusions

The spatial analysis of cities is very attractive but it is very difficult to reveal the theoretical essence of urban evolution. As Batty (2008, page 769) pointed out: "Despite a century of effort, our understanding of how cities evolve is still woefully inadequate. Recent research, however, suggests that cities are complex systems that mainly grow from the bottom up, the size and shape following well-defined scaling laws that result from intense competition for space." In order to research deeply into urban structure, we must study the scale, size, and shape of cities, and reveal the relations between them. Thus we need the concepts of scaling and dimension. The scaling relations can be directly associated with the power law, but indirectly with the exponential law (Chen and Zhou, 2008). The power law distribution and exponential distribution sometimes weave themselves together for cities. In this study, we find that the Clark model does not imply a simple exponential distribution, but suggests local and self-affine fractal form of cities. The result lends support in perspective to the suggestion of Parr (1985a, and 1985b) that the negative exponential function and the inverse power function are respectively appropriate to different urban scales.

The common exponential distribution is not fractal, but it has a global dimension in theory ($d=2$) and local dimension in practice ($D \leq 2$). By the Taylor series expansion and Fourier transform, we can derive these two dimension values, a global dimension and a local dimension, both of which equal 2 in theory. The common normal distribution can be treated and understood in the same way. Actually, the normal function and exponential function can be formally unified into a generalized exponential function. Where the dimension is concerned, the normal distribution is very similar to the exponential distribution. If the exponential distribution is based on fractal supports rather than Euclidean plane, the characteristic radius of a city will scale with the radius of study area. In this instance, the global dimension can be decomposed and replaced by two local dimensions: inner dimension and outer dimension. The global dimension cannot be directly calculated, but can be indirectly estimated by Fourier analysis. The inner dimension is still a Euclidean dimension in theory, but it can be treated as a quasi-fractal dimension in empirical analysis. The outer dimension is a fractal dimension come between 1 and 2. The normal distribution can be comprehended and dealt with through the same approach.

The exponential distribution based on the fractal background takes on self-affine fractal form. From the center to exurban region, a city can be divided into three layers. The area within the first circle is the inner layer, which is a small scale-free region with a quasi-fractal dimension $d=2$ in theory or $D<2$ in practice. The layer outside the second circle is the outer part, which is a vast scale-free space with a fractal dimension varying from 1 to 2. The area coming between the first circle and the second circle is the middle layer, which is a memory-free region and represents the conventional exponential distribution. The two circles as dividing lines should be replaced by isolines in practice. The normal distribution can be treated in the similar way. As for the power-law distribution, it can also be divided into three layers, but the nature of each part is contrary to the model of the exponential distribution.

**Acknowledgment**: This research was sponsored by the National Natural Science Foundation of China (Grant No. 41171129, 40971085). The supports are gratefully acknowledged. Many thanks to the three anonymous reviewers who provided interesting suggestions.

# Appendix

## The similarity of Hangzhou's urban form to the DLA model

The DLA model is originally proposed by Witten and Sander (1981) to simulate metal-particle aggregation. Based on Matlab, the simulation procedure of two-dimensional aggregates is as follows. (1) Define a round field on a 2-dimension Euclidean plane. (2) Set a particle at the origin as a seed. (3) Put another particle into the field, and let it move randomly. The starting point is far from the seed. If the moving particle come very close to and finally touches the seed, it will stop moving and become a part of the aggregate. Otherwise, let it vanish. Then introduce the second particle in the same way and let it walk randomly until it touch the growing cluster and form part of the aggregate or disappear due to going beyond the field's boundary. Repeat this process more than 5000 times until the aggregate comprises 5000 particles except the seed (Figure A). Fitting the cumulative number of particle within the radius of $r<64$ to equation (2) yields a cumulative exponential distribution model such as

$$P(r) = 23744.3366[1-(1+\frac{r}{82.7862})\exp(-\frac{r}{82.7862})].$$

The goodness of fit is about $R^2=0.9997$. The observed values are well consistent with the predicted values (Figure B). Apparently, the cumulative distribution function of the DLA cluster is the same as the population distribution function of Hangzhou city (see Subsection 3.1).

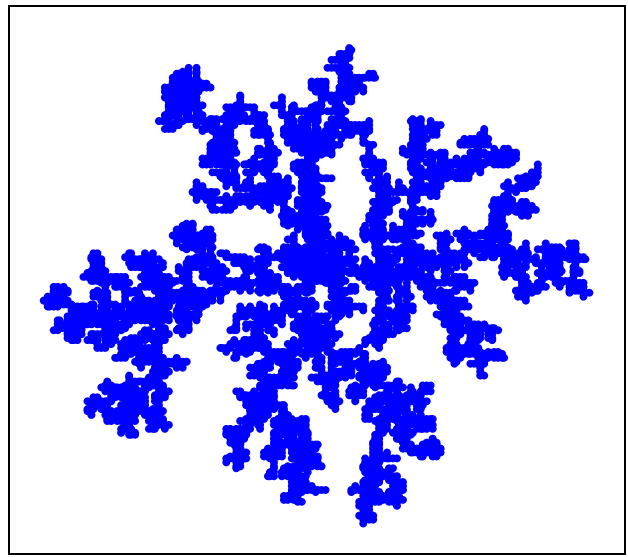

**Figure A A DLA pattern yielded by the Brownian motion with 5000 particles**

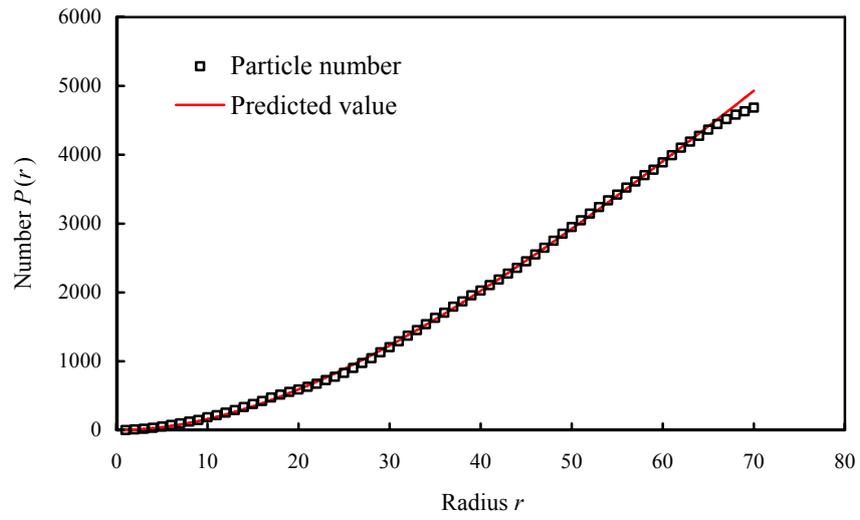

**Figure B The relation between radius and corresponding cumulative particle number of DLA**